\documentclass[%
 aip,
 amsmath,amssymb,
 reprint,%
]{revtex4-1}

\usepackage{graphicx}% Include figure files
\usepackage{dcolumn}% Align table columns on decimal point
\usepackage{bm}% bold math
\usepackage[utf8]{inputenc}
\usepackage[T1]{fontenc}
\usepackage{mathptmx}
\usepackage{siunitx}
\usepackage[normalem]{ulem}

\usepackage{hyperref}

\begin{document}

\preprint{AIP/123-QED}

\title{Noncovalent force spectroscopy using wide-field optical and diamond-based magnetic imaging}

\author{S. Lourette}
\email{slourette@berkeley.edu}
\affiliation{Department of Physics, University of California, Berkeley, California 94720-7300, USA}

\author{L. Bougas}
\affiliation{Institut für Physik, Johannes Gutenberg Universität-Mainz, 55128 Mainz, Germany}

\author{M. Kayci}
\affiliation{Department of Physics, University of California, Berkeley, California 94720-7300, USA}
\affiliation{DWI-Leibniz Institute for Interactive Materials, Aachen 52074, Germany}

\author{S. Xu}
\affiliation{Department of Chemistry, University of Houston, Houston, TX 77204, USA.}

\author{D. Budker}
\affiliation{Department of Physics, University of California, Berkeley, California 94720-7300, USA}
\affiliation{Helmholtz Institute Mainz, Johannes Gutenberg University, 55099 Mainz, Germany}

\date{\today}

\begin{abstract}
A realization of the force-induced remnant magnetization spectroscopy (FIRMS) technique of specific biomolecular binding is presented where detection is accomplished with wide-field optical and diamond-based magnetometry using an ensemble of nitrogen-vacancy (NV) color centers. The technique may be adapted for massively parallel screening of arrays of nanoscale samples.
\end{abstract}

\maketitle

\section{Introduction}

Detecting molecular targets with high specificity and sensitivity is of importance in various areas of research, ranging from drug-design applications to diagnostics and disease monitoring. Existing techniques include those based on atomic force microscopy (AFM), \cite{AFM} surface plasmon resonance (SPR), \cite{SPR} optical tweezers, \cite{tweezers} surface-enhanced Raman scattering (SERS), \cite{SERS} acoustic force spectroscopy (AFS), \cite{AFS} and force-induced remnant magnetization spectroscopy (FIRMS). \cite{FIRMS1,FIRMS2} FIRMS measures molecular bond strengths through the use of gradually increasing forces, typically applied using a centrifuge, or recently ultrasound. This is done by detecting the magnetic field produced by a collection of magnetic microspheres (beads) that label specific molecular targets. The beads are magnetized and bound to a surface by the specific interaction of interest, in the presence of a magnetic field. The magnetic field is then removed and the applied force is increased, breaking each bond as the applied force exceeds its rupture force. This results in an abrupt change in the observed magnetic field from the collection of magnetic beads corresponding to the strength of the specific interaction. \cite{FIRMS1,FIRMS2}

FIRMS, through its high force resolution, can precisely distinguish noncovalent intermolecular interactions through high force resolution, for example those of DNA duplexes with a single basepair difference. FIRMS offers high force resolution that can precisely distinguish noncovalent intermolecular interactions, for example those of DNA duplexes with a single basepair difference. \cite{FIRMS1,FIRMS2} However, FIRMS currently uses vapor-based atomic magnetometers for detection, with sensor sizes on the order of millimeters. Due to the size mismatch, the atomic magnetometers  cannot detect individual magnetic beads that are used to label the biomolecules. A more optimal sensor would detect magnetic fields on the micron scale, matching the size of the beads, and could resolve individual beads while detecting much larger magnetic fields, measuring closer to the source. We set out to improve upon FIRMS by shifting the detection to magnetic microscopy, with a goal of creating a compact and robust instrument capable of background-free, optical diffraction-limited imaging and high-throughput quantitative characterization of bond strength.

In this work, a proof-of-principle experiment is realized, combining FIRMS with magnetic and optical microscopy in order to detect biotin-streptavidin bonds, chosen for their well-characterized interaction. \cite{BiotinStreptavidin1,BiotinStreptavidin2,BiotinStreptavidin3} Streptavidin-coated magnetic microspheres are allowed to bind to biotinylated and control (non-biotin) diamond surfaces, which are then shaken with piezo-electric actuators. In this study we replace the vapor cell magnetometer with a planar ensemble of nitrogen-vacancy (NV) centers, thus boosting the detected signal by many orders of magnitude, providing optical diffraction-limited single-microsphere resolution. The biotin and control surfaces are prepared directly on the diamond plates, in order to minimize the distance between the microspheres and the NV centers and to create a compact and robust instrument.

Both magnetic imaging and wide-field optical imaging techniques are used to detect detachment events. Magnetic imaging provides background-free detection of the positions and orientations of the particles, and is available in the absence of optical access, such as when using an opaque medium. Wide-field optical imaging offers higher throughput, allowing for the tracking of moving particles. Together, these two imaging techniques provide correlated detection of individual particles and form a powerful toolset for quantitatively characterizing binding behaviors.

We observe a distinct and reproducible binding behavior when using control and functionalized surfaces, though the observed rupture forces are in a different force regime than typically observed in similar environments. Individual microbeads are magnetically imaged and subtle changes in angular orientation between force applications are resolved.
The technique can be extended to nanoscale samples, enabling massively parallel spatially resolved measurements.

\section{Methods}

\subsection{Diamond preparation}

The diamonds used in this study are electronic grade chemical-vapor-deposition (CVD) plates, produced by Element 6, roughly 100 \si{\micro\meter} in thickness. Each plate was implanted with $^{15}$N ions of energies and doses listed in the table below, with the goal of creating a $\approx$100 nm uniform high-density layer of NVs. $^{15}$N was chosen for its narrower spectrum on account of its nuclear spin of 1/2 rather than spin~1. \cite{N15}

\begin{table}[hb]
    \caption{$^{15}$N Implantation Parameters \hspace{\textwidth}}
    \label{impantation}
    \begin{center}
        \begin{tabular}{|c||c|}
            \hline
            Energy (keV) & Dose (cm$^{-2}$)\\
            \hline
            10 & $4$ $\cdot$ $10^{12}$\\
            20 & $6$ $\cdot$ $10^{12}$\\
            35 & $9$ $\cdot$ $10^{12}$\\
            60 & $1.4$ $\cdot$ $10^{13}$\\
            100 & $2.2$ $\cdot$ $10^{13}$\\
            \hline
        \end{tabular}
    \end{center}
\end{table}

After implantation, the plates were annealed \cite{AnnealRef} in an oven for 2 hours at \SI{800}{\celsius} and 4 hours at \SI{1100}{\celsius}. To prepare the surface of the diamond to bind with streptavidin-coated microparticles, the plates are functionalized with an Alkene-EG-COOH crosslinker, which bonds to biotin to form a biotinylated surface. Instead of using bare diamond, control plates were also prepared through functionalization with an Alkene-EG-COOH crosslinker, omitting the step of bonding to biotin. Diamond functionalization was performed for us by Adamas Nanotechnologies.\footnote{Adámas Nanotechnologies. \url{https://www.adamasnano.com/}}

\subsection{Imaging magnetometer}

\begin{figure}[hb]
    \centering
    \includegraphics[width=0.47\textwidth]{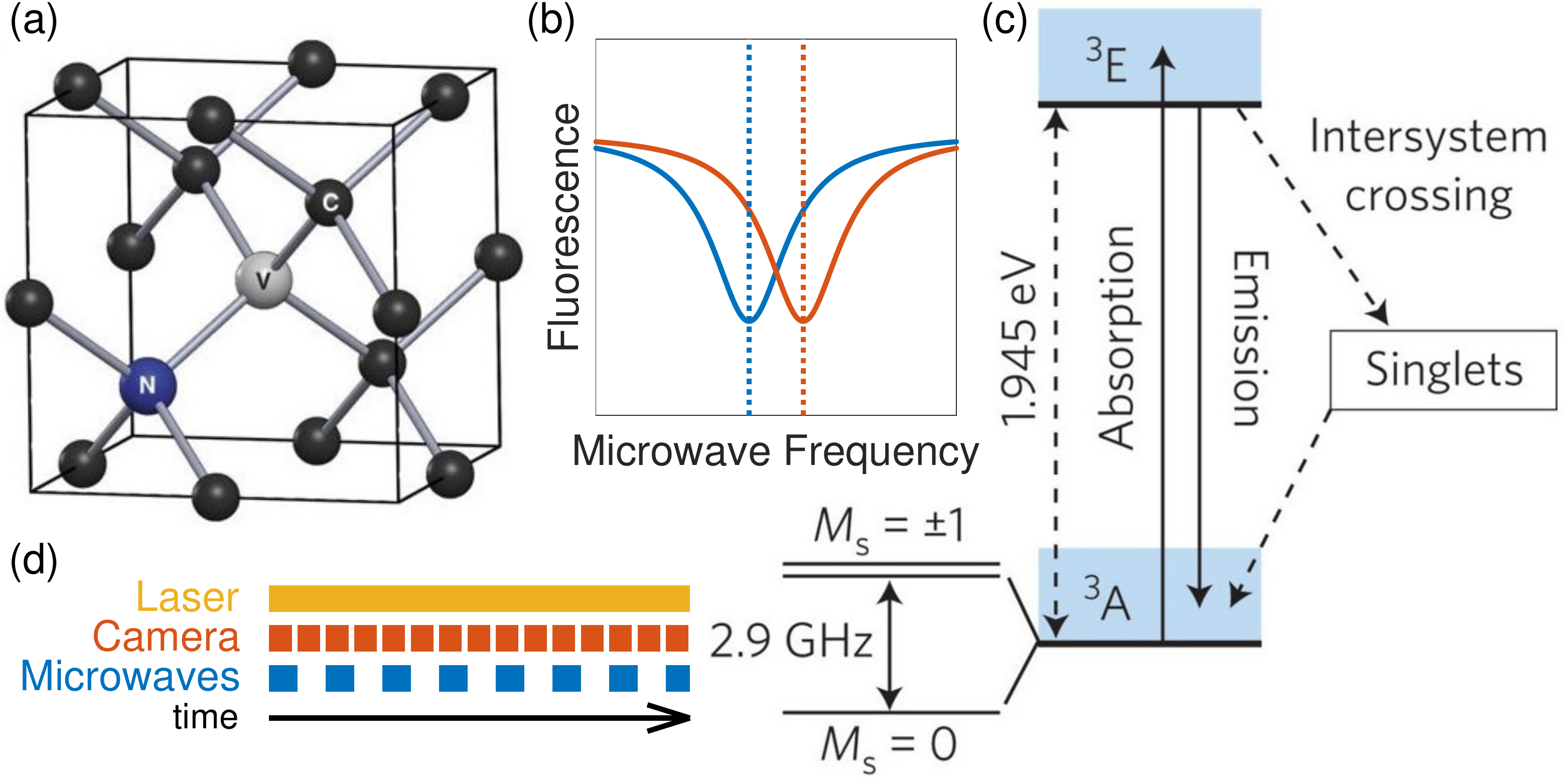}
    \caption{\label{fig:NV}Operating principle of NV-based magnetometry. (a) The NV center is composed of a substitutional replacement of a pair of adjacent carbon atoms with a nitrogen atom and a vacancy. (b) The ODMR technique can be used to measure magnetic field strength by sweeping the frequency of an applied microwave field, and observing a decrease in fluorescence that appears on resonance. (c) A spin-preserving optical transition between ground and excited triplet spin states with phonon sidebands. (d) The pulse sequence for cw magnetometry with lock-in detection consists of a camera triggering at twice the frequency of microwave modulation to ensure that microwaves are present in alternating images, while the laser remains active throughout the experiment. The images in (a) and (c) are partially based on Refs. \onlinecite{NV_Lattice,NV_Levels}.}
\end{figure}

The operating principle of the magnetometer is as follows. The NV center is composed of a substitutional replacement of a pair of adjacent carbon atoms with a nitrogen atom and a vacancy (Fig.~\ref{fig:NV}a). It has a triplet spin ground state, $m_s=\lbrace{-1,0,1\rbrace}$, where at sufficiently low external fields, the 0 sublevel has lower energy due to the electron-electron interaction (Fig.~\ref{fig:NV}c). The \mbox{+1} and \mbox{-1} sublevels are sensitive to the magnetic field, with the g-factor of $\SI{2.8}{ MHz/G}$. The magnetic field strength can be calculated by measuring the frequency of the $0\rightarrow+1$ or $0\rightarrow-1$ transitions, which is typically done by employing optically detected magnetic resonance (ODMR). \cite{ODMR} An ODMR spectrum is obtained by illuminating the NV with green light and optically detecting a reduction in fluorescence when a resonant microwave field is present (Fig.~\ref{fig:NV}b). This reduction in fluorescence occurs due to a difference in the intersystem crossing rates among the three spin sublevels. \cite{ODMR}

\begin{figure}[ht]
    \centering
    \includegraphics[width=0.48\textwidth]{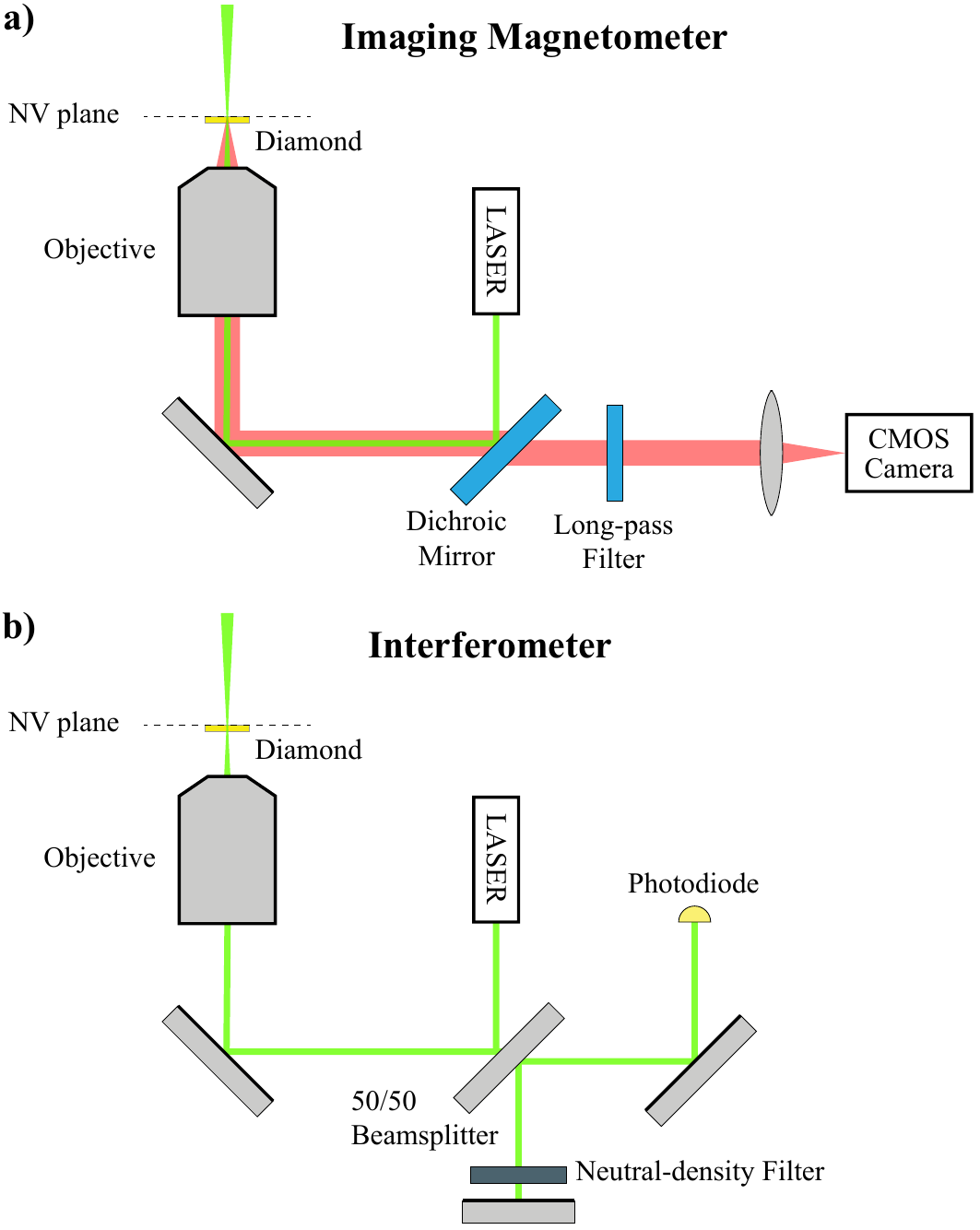}
    \caption{\label{fig:OpticalSchematic} Optical schematics for (a) imaging magnetometer and (b) interferometer. (a) Fluorescence from the layer of NV centers is collected with an infinity-corrected objective and passed through a dichroic mirror and a long-pass filter, before being imaged onto a CMOS sensor with a lens. The NV centers are excited with a beam also passing through the objective, generated by a 525 nm diode laser. The beam from the laser is made to be slightly expanding in order to increase the size of the illuminated region in the NV plane. For wide-field imaging with white light, a collimated white-light source illuminates the diamond surface directly at a normal angle of incidence, to create a bright background. (b) The dichroic mirror is replaced with a 50/50 beam splitter, and a mirror is added to form the other arm of the interferometer. A graduated neutral-density filter is inserted to balance the reflected intensity from each arm of the interferometer. The signal is detected on a fast photodiode and recorded on an oscilloscope.}
\end{figure}

A schematic of the imaging magnetometer is shown in Fig.~\ref{fig:OpticalSchematic}a. A 1 watt 525 nm diode laser (Roithner Lasertechnik NLD521000G) is used to illuminate the diamond after reflecting off of a long-pass dichroic mirror (550 nm cutoff, Thorlabs FEL0550) and passing through an objective (Olympus UPlanFl 40x). The collimation of the beam is adjusted in order to increase the size of the illuminated region at the focal plane of the objective, effectively increasing the field of view. The power of the light illuminating the sample is typically a few hundred milliwatts. The fluorescent light is collected with the objective, passes through the dichroic mirror, and is focused onto a complementary metal–oxide semiconductor (CMOS) sensor (Thorlabs DCC1240M) forming an image of the NV layer. A magnetic field of about 50 G is applied with a permanent magnet and aligned to one of the four NV axes. Using a 3D translation stage, a  loop of wire is positioned within 0.2 mm of the surface of the diamond, in order to apply microwaves to the NV centers.

The imaging magnetometer operates using a continuous wave (cw) excitation scheme: the 525 nm laser excitation is applied while the microwave frequency is swept through the resonance of the $m_s = \lbrace{ 0 \rightarrow +1 \rbrace}$ transition for the NV centers aligned along the applied magnetic field. ODMR is employed, and the resonance frequency is fitted for each pixel on the camera to form a magnetic image.

To enhance the signal quality, a lock-in detection scheme is used, in which the microwave source is switched on and off for alternating camera frames (Fig.~\ref{fig:NV}d). This is achieved by synchronizing the camera trigger to a digital trigger of half frequency, toggling between high and low for alternating images, that controls the microwave output with an RF switch. The frequency of the camera trigger is set to be as high as the camera would allow ($\approx$ 200 fps for 80x60 pixels) to avoid overexposure while maintaining the maximal exposure time.

\begin{figure*}[!ht]
    \centering
    \includegraphics[width=1\textwidth]{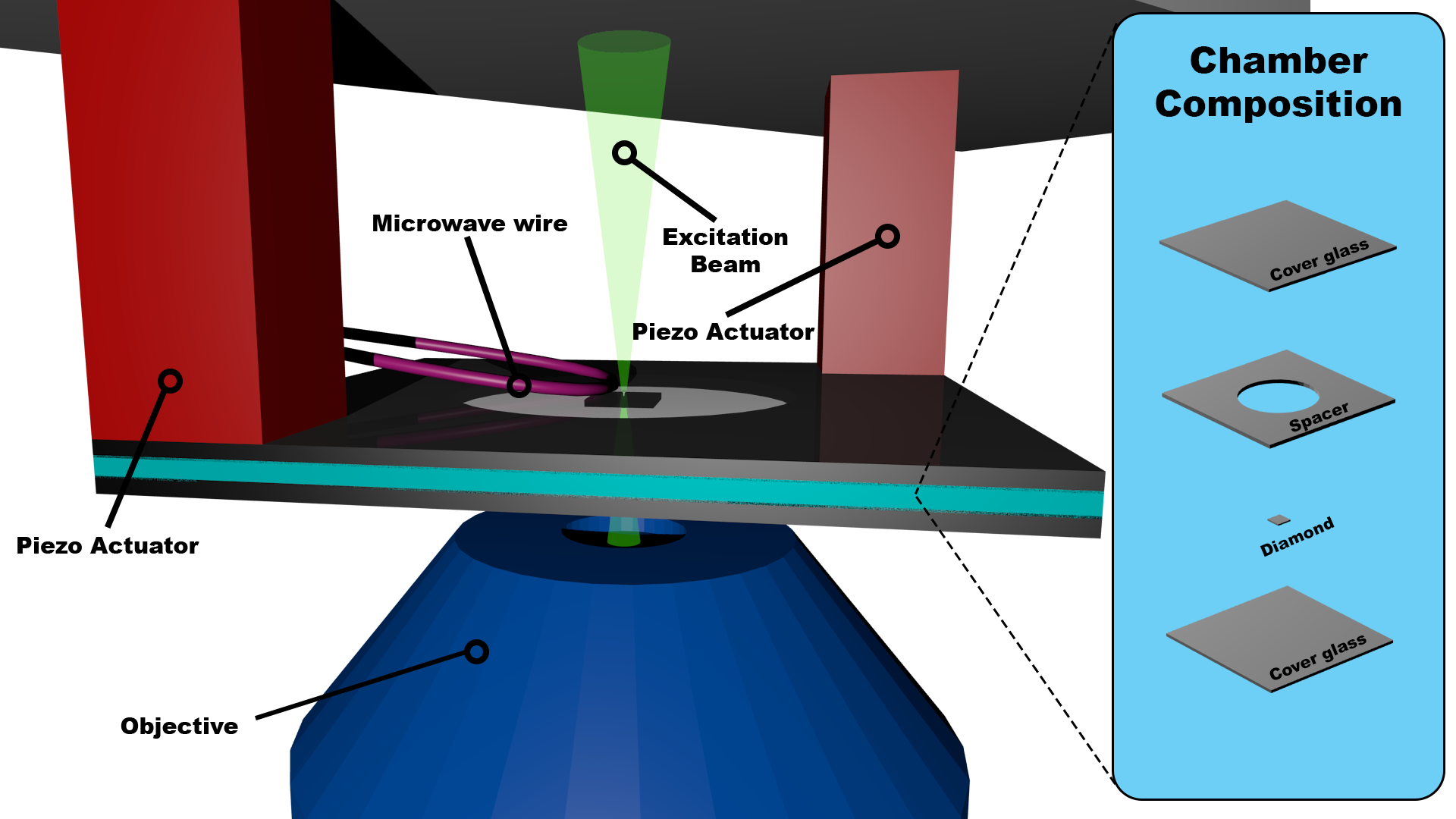}
    \caption{\label{fig:ChamberSchematic}3-D rendering of the chamber. The chamber is composed of an imaging spacer sandwiched between two glass coverslips. Before sealing the chamber, the diamond is attached to the lower coverslip with crystal wax and the chamber is filled with a solution containing the magnetic microbeads. Adhesive is used to attach a pair of piezoelectric actuators (red) to opposite corners of the chamber on one side and to a mounting plate on the other. An objective (blue), which is positioned below the chamber, delivers the excitation beam (green) to the NV centers and images the upper surface of the diamond onto a CMOS sensor. To provide the necessary microwave field to the NV centers, a copper wire (magenta) is positioned within 0.2 mm of the top surface of the chamber.}
\end{figure*}

To reduce the memory requirements of data storage and processing, instead of saving the raw frames, blocks of 32 pairs of images are summed to produce a single pair of images that are saved for later analysis. During analysis, these image pairs are converted to brightness and contrast images, where the correlated low-frequency noise cancels out in the contrast images. This process produces contrast images that are insensitive to optical excitation spatial mode and power, thermal and mechanical drift, as well as CMOS sensor inhomogeneities in both offset (dark signal non-uniformity) and gain (photo response non-uniformity) for fixed-pattern noise (FPN).

To produce a magnetic image, the microwave frequency is repeatedly swept at a rate of $\approx$ 1 MHz/s across the resonance while continuously saving contrast images. Each pixel has the frequency dependence of its contrast fitted to a double-Lorentzian with a hyperfine splitting of 3.3 MHz \cite{N15}. The fitted central frequency is converted to a magnetic field, subtracted from the background field, and combined to form a magnetic field image using the gyromagnetic ratio $\approx\SI{2.8}{MHz\per G}$.

\subsection{Microfluidics}
In this study, a sealed micro-fludic chamber is used to hold the solution containing the magnetic microspheres. The chamber is designed to be as light as possible to increase the maximum force that can be applied to the magnetic particles in-situ. A 3-D rendering of the chamber is depicted in Fig.~\ref{fig:ChamberSchematic}. The top and bottom layers of the chamber are formed by glass coverslips, and the walls of the chamber are formed by an imaging spacer (Secure-Seal\texttrademark\ Spacer from Thermo Fischer Scientific), a \SI{150}{\micro\meter} silicone gasket that is sticky on both sides. Before sealing the chamber, the diamond is attached to the center of the bottom coverslip using crystal wax with the NV side face up, and the chamber is filled with a solution of undiluted Phosphate Buffered Salene (PBS) and magnetic microbeads. The diamond is attached to the lower (rather than the upper) surface, in order to allow gravity to assist the particles in finding and bonding to the functionalized diamond surface.

\begin{figure*}%[ht!]
    \centering
    \includegraphics[width=1.0\textwidth]{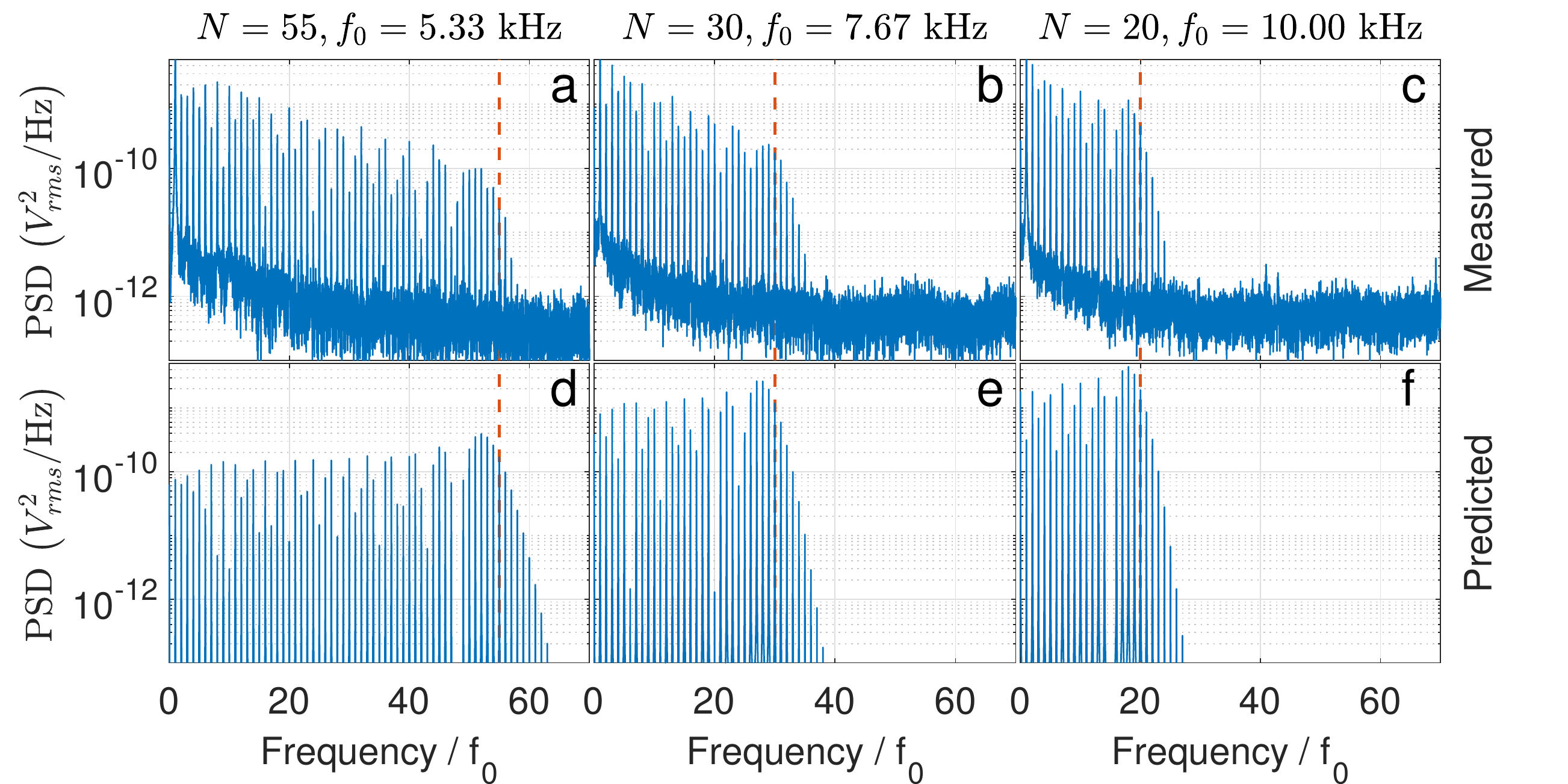}
    \caption{\label{fig:interferometersample}Power spectral density (PSD) of experimental photodiode data (a,b,c) alongside PSD of simulated data (d,e,f) with parameters chosen to yield matching spectra. The experimental data were collected using piezo drive of 60 volts peak-to-peak at frequencies of 5.33 kHz (a), 7.67 kHz (b), and 10.00 kHz (c). In each figure, the frequency axis has been scaled by $1/f_0$, such that the drive frequency and its harmonics appear at integers, allowing the value for N to be read directly from the plot (dashed vertical line). The fitted value for N can then be used to obtain a measurement of the amplitude of the mechanical motion.}
\end{figure*}

The top side of the chamber is attached to a mounting post via two piezoelectric actuator stacks (Thorlabs AE0203D08F), which attach to opposite corners of the cover glass, leaving the bottom side accessible for viewing with an objective (Olympus UPlanFl 40x). The objective is responsible for both carrying in the green light to the NVs for excitation and carrying out the red fluorescence to the camera for imaging. The wire used to couple microwaves to the NV centers is positioned as close as possible to the cover glass directly above the illuminated spot on the diamond, without itself being illuminated with the laser light. 

The device used to drive the piezo actuators is Thorlabs BPC303, a 150~V, 1~A closed-loop piezo driver. This device was used to drive the two piezo actuators in parallel with sinusoids of frequencies ranging from 5 to 10 kHz, operating close to the 1~A current limit of the device.

In order to avoid significant heating from the piezoelectric driving signal, the drive was broken up into sinusoidal pulses of 5\,ms duration, applied once per second with fixed frequency and varying amplitude. The amplitude was incremented after every 60 pulses from 0 $V_{p-p}$ to 50 $V_{p-p}$ in steps of 2 $V_{p-p}$.

The magnetic microparticles used in this study are ferromagnetic streptavidin-coated microspheres (SVFM-20) from Spherotech. The microspheres are composed of polystyrene cores and $\text{CrO}_{\text{2}}$ coatings, with an average density of 1.8 \si[per-mode=symbol]{\gram\per\centi\meter\cubed} and diameters distributed with mean, median, and standard deviation of 2.10 \si{\micro\meter}, 1.95 \si{\micro\meter}, and 0.43 \si{\micro\meter}, respectively. A particle with diameter of \SI{2}{\micro\meter} and density of \SI[per-mode=symbol]{1.8}{\gram\per\centi\meter\cubed} has an effective weight (weigh minus the buoyant force)  of \SI{3.2e-14}{\N} in PBS. Based on the distribution of sizes, one would expect 68\% of the particles to have an effective weight between \SI{1.6e-14}{\N} and \SI{5.8e-14}{\N}.

\subsection{Interferometer}

The strength of the force applied with the piezoelectric actuators can be varied by changing the drive voltage.
A force is applied to a particle by using its own inertia to pull on the bonds at the apex of its mechanical motion. As such, the applied force should be proportional to the amplitude of the mechanical oscillation of the system, which in turn we expect to be linear in applied voltage. The amplitude of the oscillations of the mechanical system was measured with an interferometer, allowing the linear response to be verified, and the linear coefficient to be obtained. With the measured amplitude and the known drive frequency, a force per unit mass, or g-force, can be calculated. This allows us to map the applied force as a function of frequency and voltage applied with the piezoelectric actuators.

A Michelson-Morley interferometer was setup by swapping the dichroic mirror for a 50/50 beam splitter, adding a mirror to form the second arm of the interferometer, and inserting a graduated neutral-density filter to balance the reflected intensity from each arm of the interferometer (Fig.~\ref{fig:OpticalSchematic}b). The output of the interferometer is focused onto a photodiode, whose signal is amplified with a current amplifier (SRS570), while maintaining a bandwidth of 200 kHz, and subsequently recorded with an oscilloscope.

In an interferometer, the two interfering beams produce an electric field and intensity at the output that can be written as \cite{Interference}
\begin{align}
    E(t) &= Re\lbrace{E_1 e^{i \omega t} + E_2 e^{i (\omega t + \phi)}\rbrace} \nonumber \\
    &= [E_1 + E_2\cos(\phi)] \cos(\omega t) + [E_2\sin(\phi)] \sin(\omega t), \nonumber \\
    \nonumber \\
    I(t) &= \langle{E(t)}^2\rangle \nonumber \\
    &=\frac{1}{2}[E_1^2 + E_2^2 + 2E_1E_2\cos(\phi)]
\end{align}
where $E_1$ and $E_2$ are the two interfering fields, $\omega$ is their angular frequency, and $\phi$ is the phase difference between the two arms of the interferometer. Furthermore,  the phase difference $\phi$ can be broken into two parts: the ideally constant piezo-independent phase distance $\phi_0$, and a term governed by the mechanical response to the piezo drive
\begin{equation}
    \phi = \phi_0 + \frac{2\pi x_0}{\lambda}\cos(2 \pi f_0 t),
\end{equation}
where $x_0$ and $f_0$ are the amplitude and frequency of the motion of the system, and $\lambda$ is the wavelength of the light. After removing the DC component, the predicted detected oscilloscope voltage signal, proportional to intensity, will be of the form
\begin{equation}
    V(t) = V_0\cos(\phi_0 + N\cos(2 \pi f_0 t)). \label{eq:interferometer}
\end{equation}
Here $N$, which we have defined to be $\frac{2\pi x_0}{\lambda}$, can be estimated by counting the number of harmonics of $f_0$ before roll-off in the power spectral density (PSD) of $V(t)$. In Fig.~\ref{fig:interferometersample}, The PSD of sample interferometer data (a,b,c) is displayed alongside the PSD of Eq.~(\ref{eq:interferometer}) (d,e,f), with matching values for $N$ and $f_0$.

\section{Results}

\begin{figure}%[!b]
    \centering
    \includegraphics[width=0.48\textwidth]{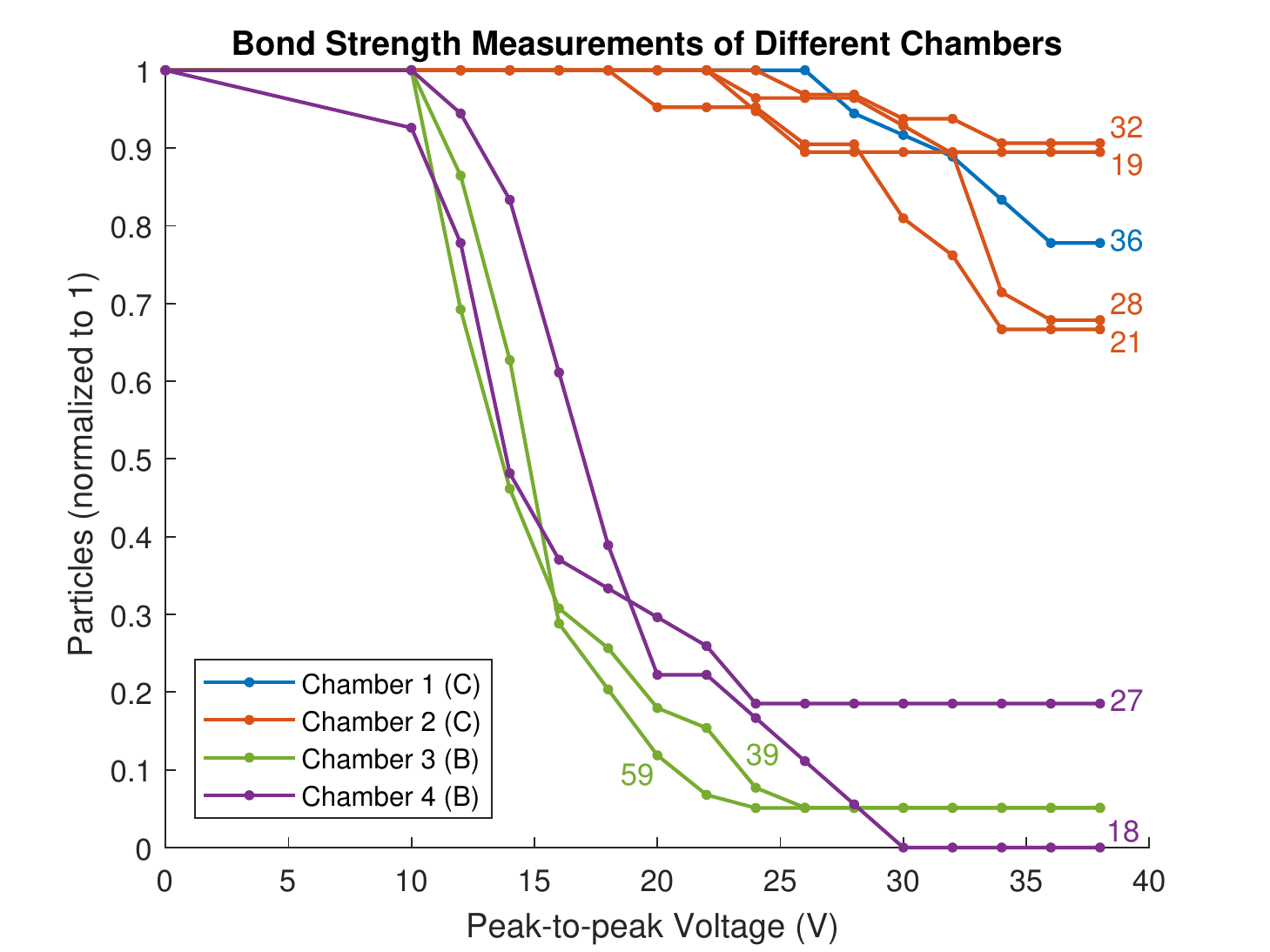}
    \caption{Normalized number of particles detached under increasingly larger voltages for various chambers. Particle positions were tracked under wide-field illumination as the voltage applied to the piezoelectric actuators was linearly ramped. The fraction of particles that remain bound to the surface is plotted against the drive voltage. The number of particles tracked during each experiment is indicated by a number labeling its corresponding line on the plot. Nine experiments were performed spanning four different chambers. Chambers with biotin surfaces are labelled with (B) and chambers with control surfaces are labelled with (C).}
    \label{fig:Reproduce}
\end{figure}

Using wide-field imaging with white light (microscope illuminator), the positions of the particles were tracked as the force was applied, and the time at which each particle detached from the surface was obtained, in order to create a plot of number of bound particles as a function of applied voltage (Fig.~\ref{fig:Reproduce}). Wide-field imaging with white light was preferred here, as it allowed for the simultaneous monitoring of dozens of particles. Reattaching particles were not counted; only particles that started attached to the surface were tracked. Particles were tracked manually, assisted by software written to facilitate this process. The results were reproducible when repeated for the same chamber; however, when a chamber was broken and a new one was assembled, results often deviated significantly.

\begin{figure}%[h]
    \centering
    \includegraphics[width=0.48\textwidth]{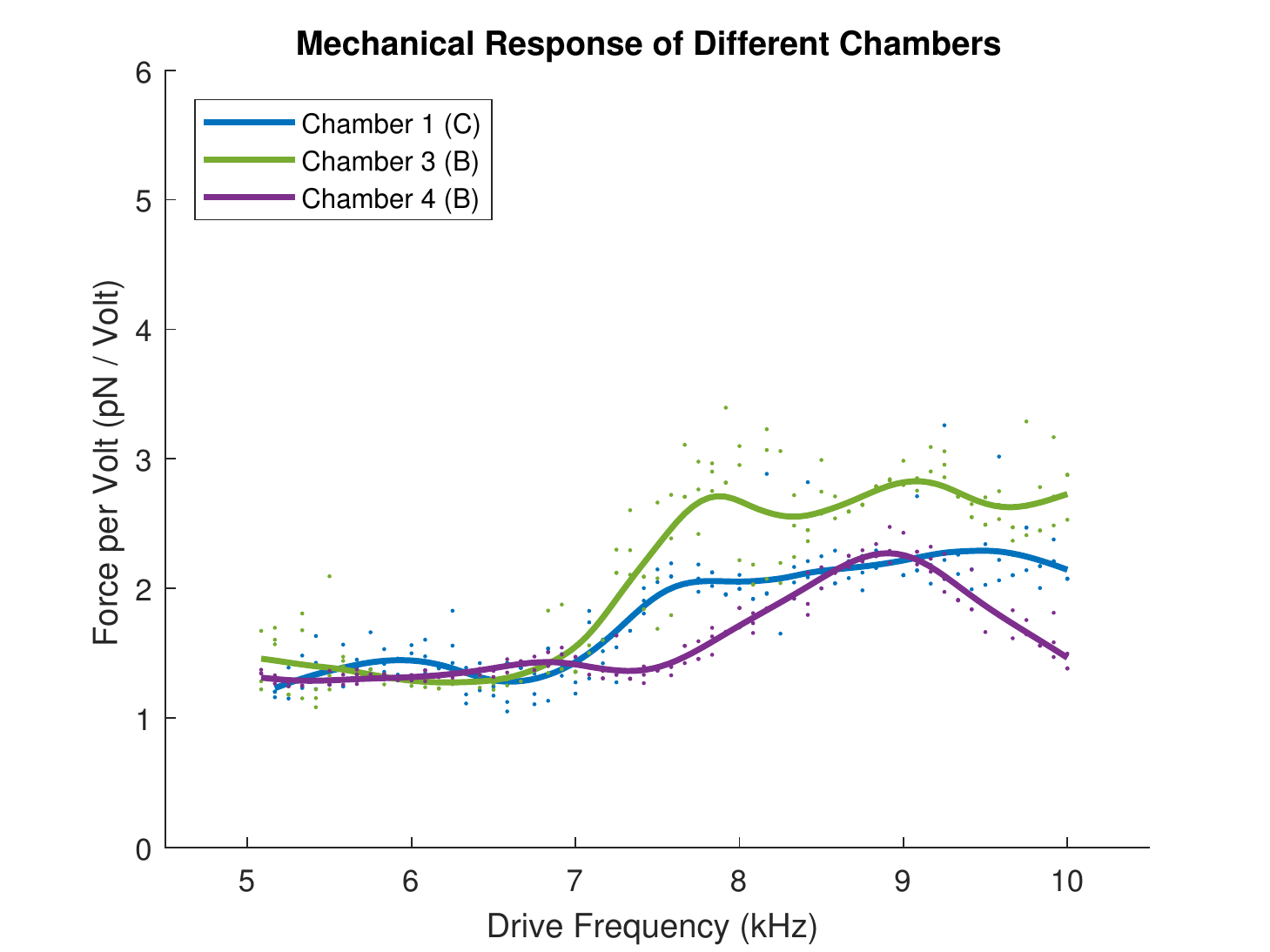}
    \caption{Amplitude of the force experienced by a microsphere with diameter \SI{2}{\micro\meter} as a function of drive frequency for different chambers.
    The force amplitude~(F) is calculated based on the sinusoidal amplitude of the mechanical motion~($x_0$), drive frequency~($\omega/2\pi$), and mass~($m$) according to the equation \mbox{$F = m \omega^2 x_0$}. Smoothing lines are added to highlight trends. Chambers with biotin surfaces are labelled with (B) and chambers with control surfaces are labelled with (C).}
    \label{fig:Response}
\end{figure}

\begin{figure*}[!ht]
    \centering
    \includegraphics[width=1.0\textwidth]{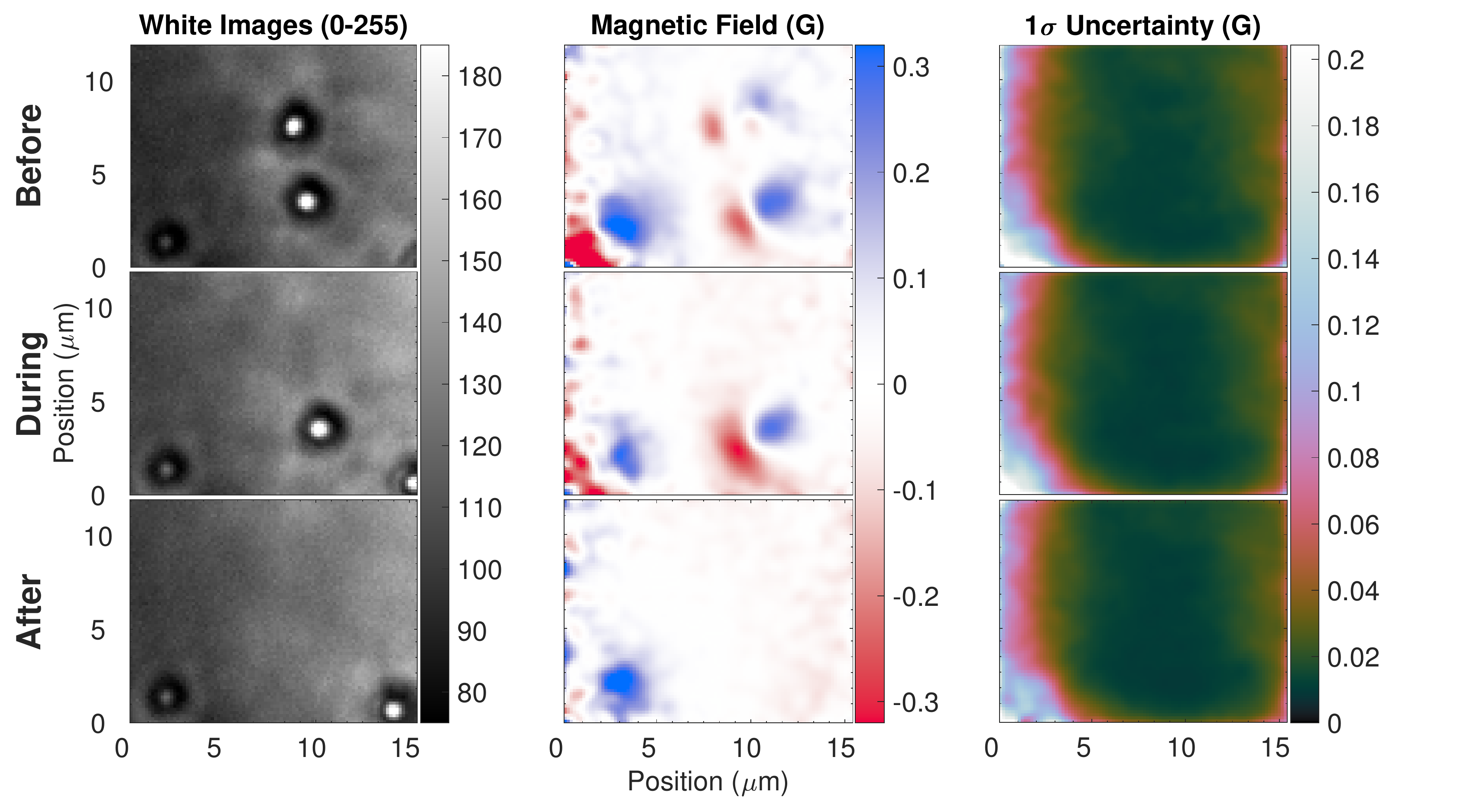}
    \caption{\label{fig:MagImgs} Images of particles before (top), during (middle), and after (bottom) application of force. Wide-field optical images are displayed on the left, where particles are visible as dark regions with Arago (Poisson) spots. Magnetic images are displayed in the middle with corresponding $1\sigma$ uncertainty displayed on the right.}
\end{figure*}

After obtaining the particle statistics from the wide-field imaging with white light, the setup was switched to the interferometer configuration and the amplitude of the motion of the chamber was measured. Using the interferometer, several chambers were analyzed to determine the applied force for various voltage drives. In Fig.~\ref{fig:Response}, the measured force experienced by a particle of diameter \SI{2}{\micro\meter} per unit of drive voltage is plotted against the drive frequency for different chambers. The plot depicts variation in resonance features between chambers, with the force at a given drive frequency and amplitude varying among chambers by anywhere from 5\% to 40\%, depending on the frequency.

The strength of  biotin-streptavidin bonds has been demonstrated to depend heavily on the loading rate,\cite{EvansRitchie1997} the rate at which the applied force is increased. It should be noted that our force application produces an oscillating force with an amplitude that is slowly ramped, rather than a constant force that is slowly ramped as is prevalent in the literature. During an oscillating force, the bond is only under stress for a fraction of each oscillation, corresponding to the duration of the time when the force is near its maximum. When taking this factor into account, we estimate an ``effective loading rate,'' or a loading rate that would be comparable to loading rates of non-alternating force applications, of $\approx\SI{70}{pN/s}$ for our experiments.

Our results indicated reproducible differences between the biotin-coated surfaces and the control surfaces, in that the former has a feature at a low rupture force. However, this force value, estimated at 20(8) pN, is a factor of 2-3 lower than literature values for rupture force of specific biotin-streptavidin, when compared to the rupture force at a similar loading rate.\cite{BiotinStreptavidin1,BiotinStreptavidin2,BiotinStreptavidin3,EnergyLandscape1999} The force range of the current setup will need to be improved to study and resolve non-specific interactions. At the current stage, it is clear that our new detection system can distinguish different surface properties.

In a separate experimental sequence, magnetic imaging data were obtained. The excitation laser was positioned over three bound particles, and the magnetic field image was obtained before, during, and after applying the pulses (described previously) to the piezoelectric actuator (Fig.~\ref{fig:MagImgs}). The pulses were paused at 40 $V_{p-p}$ in order to obtain the data for the ``during" image, and the ``after" image was obtained after ramping to 50 $V_{p-p}$. In the magnetic images, blue and red regions corresponding to positive and negative local magnetic fields with respect to applied external magnetic field, are clearly visible for each of the three particles, indicating the direction of the magnetic dipole of each particle. The maximum measured field strength is about $\SI{0.3}{G} \left(\SI{3e-5}{T}\right)$, nine orders of magnitude larger than that typically detected when performing FIRMS using a vapor cell magnetometer $\left(\SI{3e-14}{T}\right)$.\cite{FIRMS1,FIRMS_Bfield} The magnetic images have highest uncertainty along the edges of the images where the laser intensity was weakest.

\section{Discussion}

There are several factors that may introduce variance in our observations. As previously discussed, the microparticles used in these experiments have a broad size distribution (coefficient of variance of 0.2), which corresponds to a larger distribution of applied forces. Such a distribution is expected to result in an increase in the uncertainty of the measured rupture force.

It is also possible that the mechanical motion of the chamber may not be accurately predicting the applied force. Compared to the rupture force, the variation in the conversion factor (\si{pN\per V}, Fig.~\ref{fig:Response}) between applied voltage and applied force was found to be significantly smaller, at 10\% - 30\%. This leads us to believe that although the motion of the different chambers is similar, the force experienced by the particles is not.

A possible cause of the variation in the apparent force comes from a violation of the assumption that the fluid in the chamber is incompressible. Within 10 or 20 minutes of the sealing of the chamber, air bubbles can be observed along the circular edge of the chamber. These bubbles could allow for the chamber to compress and expand with each oscillation, altering the force experienced by the particles. An unsealed open chamber would not be a good alternative due to evaporation during the course of the experiment. However, a microfluidic chamber with flow appears to be a promising route for future experiment, offering a stable configuration with the benefit of reusability. We have also recently demonstrated success with a local manipulation of particles using dielectrophoresis, \cite{Metin} which appears to be a promising method for future work.

Magnetic imaging of the particles is an attractive alternative to optical wide-field imaging in the presence of significant optical backgrounds or when optical access to the sample is unavailable. The latter could be the case, for example, when using blood. Magnetic labeling has been demonstrated \cite{ForceReview,MagneticParticleReview,review_Tang} to be a valuable tool for tracking biological entities. An illustration of discriminating power of magnetic imaging can be seen in the ``during'' and ``after'' images of Fig.~\ref{fig:MagImgs}: the particle at the bottom right is apparently non-magnetic, which is impossible to tell from the optical image. Furthermore, magnetic images offer information about the orientation of the particles, which can be used to discern whether a particle has remained attached, or has detached and quickly reattached to the surface. For example, when comparing the white images of the particle in the middle of the ``before'' and ``during'' images shown in Fig. \ref{fig:MagImgs}, the particle appears to move slightly to the right, but it is not known whether it is from sliding or rolling across the surface. When we examine the corresponding magnetic images and fit the orientation, we observe a spatial rotation with a total angle of 10(2) degrees.

When compared against the original FIRMS technique, this method offers significantly (by nine orders of magnitude) boosted signal strength, allowing for the resolution of individual particles, in addition to in-situ force application. This technique also allows for massively parallel experiments with various particles and/or surface functionalizations in a lab-on-a-chip environment. However, an important difference is that the field-of-view of the current microscopy-based approaches is much smaller, by two to three orders of magnitude, than those of FIRMS using an atomic magnetometer. Consequently, a much smaller number of molecular bonds are measured, which may affect the statistics of the force distribution.

\section{Conclusion}

We successfully realized a diamond-based FIRMS experiment capable of resolving individual particles with both optical and magnetic imaging, in addition to quantifying the strength of the molecular interactions. Although we found that with our method of applying force, the apparent rupture force of the particles varied significantly between chambers, this can potentially be addressed with an alternate method of force applications, such as dielectrophoresis, as previously demonstrated in our laboratory. \cite{Metin} This technique may prove useful in drug testing and diagnostics: we envision a system of multiple parallel microfluidic channels carrying nanoparticles with various functionalizations that are detected with high sensitivity, and whose targeted molecular interaction is quantitatively measured.

\section{Acknowledgements}
The authors acknowledge technical help and advice from Tao Wang, Andrey Jarmola, as well outstanding support with diamond surface functionalization from Olga Shenderova, and Marco Torelli of Adámas Nanotechnologies. This work was supported in part by the EU FET OPEN Flagship Project ASTERIQS. SX acknowledges partial support by the NSF (1508845).

\nocite{*}
\bibliography{D-FIRMS}% Produces the bibliography via BibTeX.

\end{document}